\documentclass[accepted]{article}

\usepackage{microtype}
\usepackage{graphicx}
\usepackage{subfigure}
\usepackage{booktabs} 

\usepackage{hyperref}

\usepackage{icml2022}

\usepackage{amsmath}
\usepackage{amssymb}
\usepackage{mathtools}
\usepackage{amsthm}

\usepackage[capitalize,noabbrev]{cleveref}

\theoremstyle{plain}

\theoremstyle{definition}

\theoremstyle{remark}

\usepackage[textsize=tiny]{todonotes}

\icmltitlerunning{Burst2Vec}

\usepackage{microtype}
\begin{document}

\twocolumn[

\icmltitle{Burst2Vec: An Adversarial Multi-Task Approach for \\Predicting Emotion, Age, and Origin from Vocal Bursts}



\icmlsetsymbol{equal}{*}

\begin{icmlauthorlist}
\icmlauthor{Atijit Anuchitanukul}{contex}
\icmlauthor{Lucia Specia}{contex}
\end{icmlauthorlist}

\icmlaffiliation{contex}{Contex.ai, London, United Kingdom}

\icmlcorrespondingauthor{Atijit Anuchitanukul}{atijit.anuchitanukul20@imperial.ac.uk}

\icmlkeywords{Machine Learning, ICML}

\vskip 0.3in
]



\printAffiliationsAndNotice{}  

\begin{abstract}
We present Burst2Vec, our multi-task learning approach to predict emotion, age, and origin (i.\,e., native country/language) from vocal bursts. Burst2Vec utilises pre-trained speech representations to capture acoustic information from raw waveforms and incorporates the concept of model debiasing via adversarial training. Our models achieve a relative 30\,\% performance gain 
over baselines using pre-extracted features and score the highest amongst all participants in the ICML ExVo 2022 Multi-Task Challenge.

\end{abstract}

\section{Introduction}
\label{introduction}
Substantial research has been dedicated to analysing and predicting various outputs from speech data beyond speech recognition. For example, detecting emotion~\cite{TZIRAKIS202146}, age~\cite{schuller2010compare}, cognitive load~\cite{schuller2014interspeech}, and COVID-19~\cite{schuller2021compare}.
Much less studied~\cite{vol1, vol2} is the problem of predicting these and other signals from {\em vocal bursts}, i.\,e., non-linguistic vocalisations such as laughs, gasps, and cries, which are key to the expression of emotion and human communication in general~\cite{BairdExVo2022}.

The ICML ExVo Challenge~\cite{BairdExVo2022}\footnote{\url{https://www.competitions.hume.ai/exvo2022}} introduced the first large-scale dataset of human vocalisations as well as baseline approaches to enable research in computational approaches for understanding and generating vocal bursts. In this paper, we focus on understanding vocal bursts as part of the Multi-Task Learning Track of the Challenge. This track requires participants to train a single model to jointly recognise ten types of emotions and two demographic traits from vocal bursts: age and country of origin.


We propose a novel approach to this task which is based on two main components: (1) transfer-learning from strong pre-trained speech representations; (2) multi-task learning with an adversarial training strategy to better debias the model by disentangling task-specific from general representations. 


We explore transfer learning from two popular self-supervised models for speech representation learning, namely 
wav2vec 2.0~\cite{wav2vec2} and HuBERT~\cite{hubert}. Although these models have been originally proposed for speech recognition, they have also been successfully applied in studies for emotion recognition due to their ability to capture prosodic information and emotion-dependent word patterns~\cite{9383456, pepino21_interspeech, 9747417}. We investigate the effect of these pre-trained models in our non-verbal communication tasks, which we believe has not been explored before.


Multi-Task Learning (MTL) has been extensively explored 
to leverage shared  information across related tasks. 
However, depending on the settings, MTL training can bias the model by giving more importance to representations from certain tasks~\cite{Ruder2017AnOO}, which  can hurt  performance on other tasks. 
Our MTL approach differs from existing work in that adversarial training is applied with the focus of disentangling each input representation into a shared representation (with information for all tasks), and task-specific representations (with information  only relevant for a given task). We show that this is beneficial when unintended biases are present in the dataset between tasks.

\section{Dataset and Analysis}
\label{subsec:data_analysis}

\begin{figure}[!h]
\centerline{\includegraphics[scale=0.5]{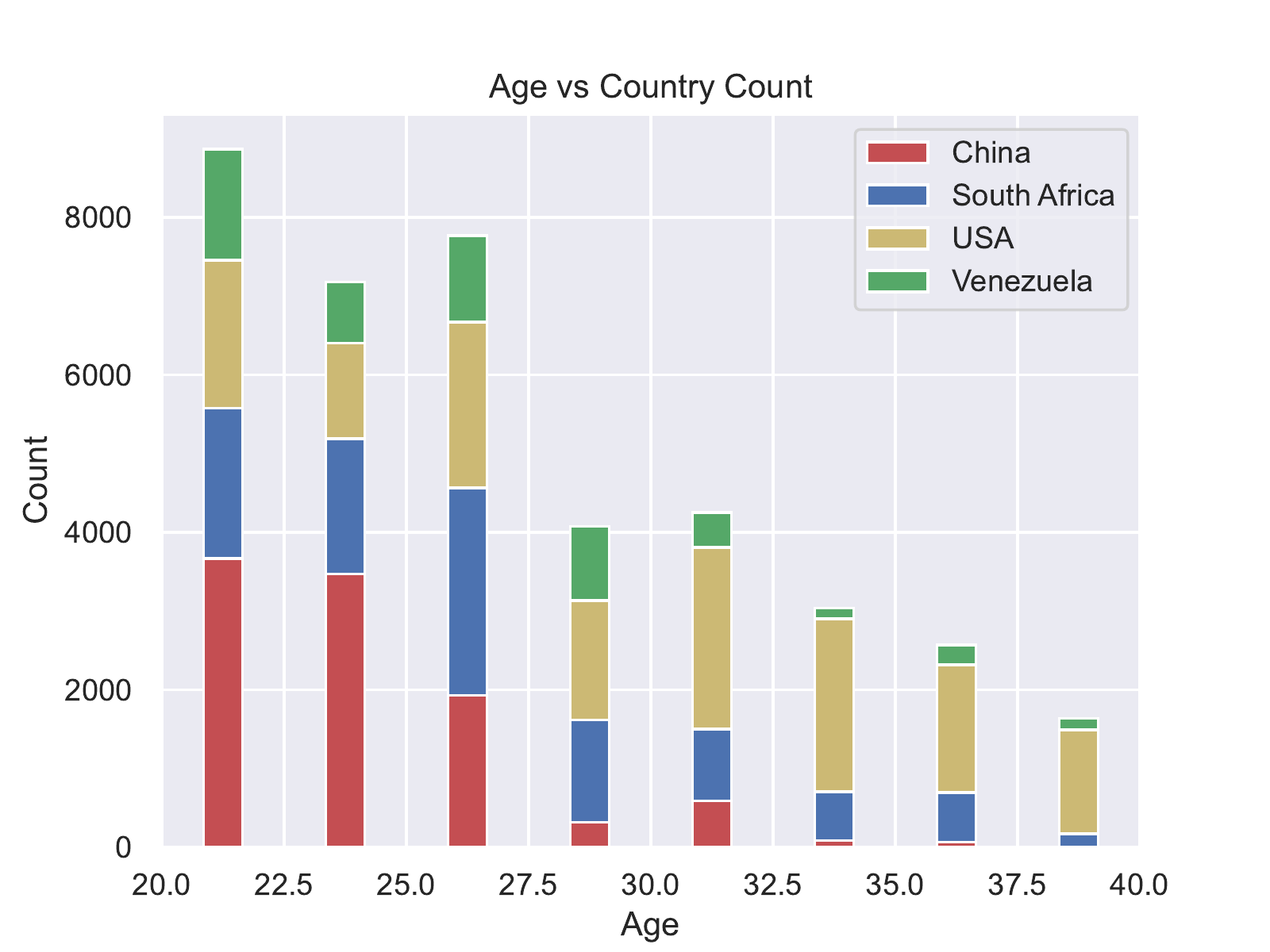}}
\caption{Distribution of age and country labels for the combined training and validation sets.}
\label{fig:age_vs_country}
\end{figure}

The dataset contains over $36$ hours of vocal bursts from $1,702$ speakers from China, South Africa, USA, and Venezuela (Table \ref{Tab:data_overview}). The audio files were collected `in-the-wild' by the speakers. Each vocal burst is self-annotated
by their speakers with intensities in [1:100] of ten different emotions, \emph{Amusement}, \emph{Awe}, \emph{Awkwardness}, \emph{Distress}, \emph{Excitement}, \emph{Fear}, \emph{Horror}, \emph{Sadness}, \emph{Surprise}, and \emph{Triumph}. The age range is from 20 to 39\,years. 
\begin{table}[h!]
   \centering
   \footnotesize
   \scalebox{0.9}{\begin{tabular}{l|rrrr}
        \hline\hline
        & \textbf{Train} & \textbf{Validation} & \textbf{Test}\\
        \hline
        \textbf{HH:MM:SS} & 12:19:06 & 12:05:45 & 12:22:12\\
        \textbf{No.} & 19990 & 19396 & 19815\\
        \hline
        \textbf{Speakers} & 571 & 568 & 563\\
        \textbf{F:M} & 305:266 & 324:244 & -\\
        \textbf{USA} & 206 & 206 & -\\
        \textbf{China} & 79 & 76 & -\\
        \textbf{South Africa} & 244 & 244 & -\\
        \textbf{Venezuela} & 42 & 42 & -\\
        \hline\hline
    \end{tabular}}
  \caption{Statistics of the ExVo Challenge dataset including the duration (HH:MM:SS), sample count (No.), speaker count, gender ratio (F:M), and per-country speaker counts~\cite{BairdExVo2022}.}
  \label{Tab:data_overview}
  \vspace{-10pt}
\end{table}
Upon analysis, we found that the age and country distribution is imbalanced, as indicated in Figure~\ref{fig:age_vs_country}. In particular, there are more participants younger than 27.5\, years and participants from China, South Africa, and Venezuela are younger in general, whereas American participants are older. This can bring an unintended cross-label bias to MTL approaches.

\section{Methodology: Burst2Vec}
\subsection{Architecture Overview}
\begin{figure}[!h]
\centerline{\includegraphics[scale=0.5]{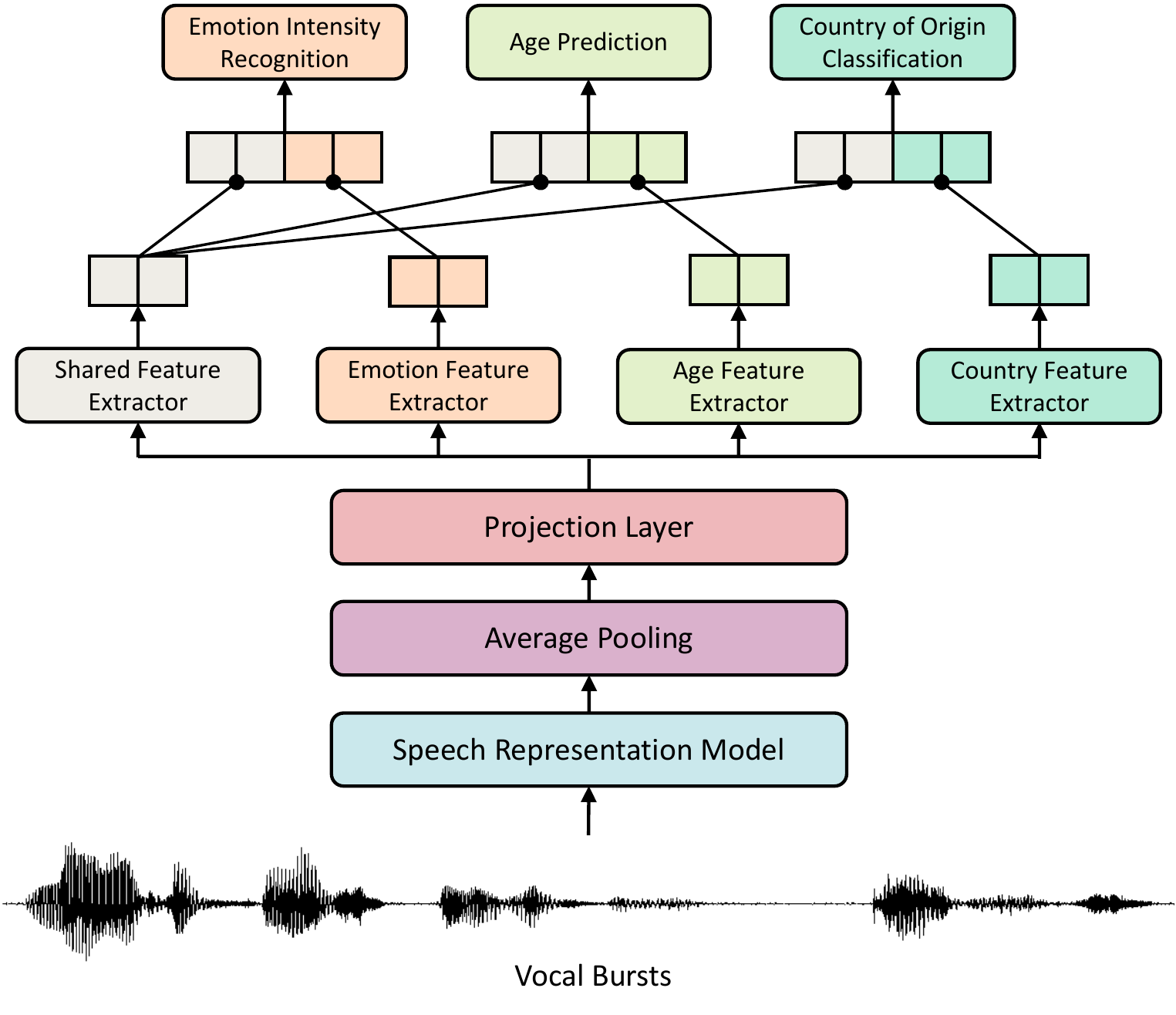}}
\caption{Overview of the Burst2Vec architecture.}
\label{fig:burst2vec}
\vspace{-10pt}
\end{figure}

We propose Burst2Vec, which is shown in Figure~\ref{fig:burst2vec}. It takes in raw audio waveform of vocal bursts as input. Then, a speech representation model (\S\ref{subsec:speech_repr}) is used to compute the temporal acoustic representations. After applying average pooling to obtain vector representations of the bursts, a projection layer projects the pooled representations to a lower dimensional space.

Following \cite{Li_MTL:22}, the projected representations are then passed to two feature extractors for shared and task-specific features. With adversarial debiasing (\S\ref{subsec:adv_debias}), the shared feature extractor is trained to extract features useful for the three tasks, while each task-specific feature extractor captures features that are only beneficial to the corresponding task. This is to prevent the model from learning spurious relationships between tasks due to biases in the dataset such as the one mentioned in \S\ref{subsec:data_analysis}. At inference, the shared and task-specific representations are concatenated and passed through a projection layer to form combined representations for the corresponding output head. 

\subsection{Pre-trained Speech Representation} \label{subsec:speech_repr}
We explore two pre-trained speech representation models: wav2vec 2.0~\cite{wav2vec2} and Hidden-unit 
BERT 
(HuBERT)~\cite{hubert}. We posit that these  models are able to capture not only linguistic, but also general acoustic information, and therefore can be beneficial for this task. 
Also, given the proven benefits of transfer learning from large pre-trained models, we expect that fine-tuning on such models will lead to better results than relying on pre-extracted features, even though those have been designed specifically to capture acoustic information.

Both models are trained on speech data in a self-supervised manner with a Transformer architecture. See Appendices~\ref{sec:wav2vec2_description} 
and~\ref{sec:hubert_description} for details on  wav2vec 2.0 and HuBERT.

\subsection{Tasks and Loss Functions}

\paragraph{Emotion Recognition} Different from the approch released by the task organisers, our loss function is based on the Concordance Correlation Coefficient (CCC) metric (Eq.\ 1). The resulting loss ($\mathcal{L}_{CCC}$) is defined in Eq.\ 2.
\begin{gather}
    \vspace{-25pt}
    CCC = \frac{1}{10}\sum_{i=1}^{10}\frac{2\sigma_{\mathbf{x}_{i}\mathbf{y}_{i}}^{2}}{\sigma_{\mathbf{x}_{i}}^{2} + \sigma_{\mathbf{y}_{i}}^{2} + (\mu_{\mathbf{x}_{i}} - \mu_{\mathbf{y}_{i}})^2}\\
    \mathcal{L}_{CCC} = 1 - CCC, 
    \vspace{-5pt}
\end{gather}
where $\mathbf{x}_{i}$ and $\mathbf{y}_{i}$ are the predictions and the ground truth, respectively, for each of the ten emotions, while $\mu_{\mathbf{x}_{i}} = \mathbb{E}(\mathbf{x}_i)$, $\mu_{\mathbf{y}_{i}} = \mathbb{E}(\mathbf{y}_i)$, $\sigma_{\mathbf{x}_{i}}^2 = \text{var}(\mathbf{x}_i)$, $\sigma_{\mathbf{y}_{i}}^2 = \text{var}(\mathbf{y}_i)$, and $\sigma_{\mathbf{x}_{i}\mathbf{y}_{i}}^2 = \text{cov}(\mathbf{x}_i, \mathbf{y}_i)$.

\paragraph{Age Prediction} We employ the Mean Absolute Error loss function ($\mathcal{L}_{MAE}$), which is defined as follows:
\begin{gather}
    \mathcal{L}_{MAE} = \frac{1}{N}\sum_{n=1}^{N}|x_n - y_n|,
\end{gather}
where $x_n$ and $y_n$ are the predicted age and target age, respectively, for an audio input $n$.

\paragraph{Country of Origin Classification} For this task, we employ the Cross-Entropy loss function.
\begin{gather}
\mathcal{L}_{CE} = -\frac{1}{N}\sum_{n=1}^{N}\log\frac{\exp(x_{n,y_{n}})}{\sum_{c=1}^{4}\exp(x_{n,c})},
\end{gather}
where $x_{n,c}$ is the output logit from the model of the audio input $n$ corresponding to the country class $c$.

\paragraph{Combined Output Loss} During training, to ensure the individual effectiveness of the shared and task-specific representations, each output head performs three forward passes with shared representation ($\mathbf{r}_{Shared}$), task-specific representation ($\mathbf{r}_{Spec}$), and projected concatenation of shared and task-specific representations ($\mathbf{r}_{Concat}$) passed as separate inputs. Thus, for task $j$, the combined output loss ($\mathcal{L}_{O}^j$) is the sum of losses from the three forward passes.
\begin{gather}
    \mathcal{L}_{O}^j = \mathcal{L}_{Shared}^j + \mathcal{L}_{Spec}^j + \mathcal{L}_{Concat}^j.
\end{gather}

\subsection{Adversarial Debiasing} \label{subsec:adv_debias}
\paragraph{Task Discrimination Loss} Following \cite{Li_MTL:22}, to ensure that the shared feature extractor only captures features that are beneficial across all tasks, we perform adversarial training between the shared extractor and a task discriminator. Given a batch of task $j$, the discriminator loss ($\mathcal{L}_{D}^j$) is defined as follows:
\begin{gather}
    \mathcal{L}_{D}^j = -\frac{1}{N}\sum_{n=1}^{N}\log\frac{\exp(x_{n,j})}{\sum_{t\in\mathbb{T}}\exp(x_{n,t})},
\end{gather}
where $\mathbb{T} = \{Emotion, Age, Country\}$, and $x_{n,t}$ is the output logit from the discriminator of the audio input $n$ corresponding to task $t$. To update the weights of the shared feature extractor, the gradient of the discriminator loss is reversed via the gradient reversal layer so that the weights are updated in the ascending direction along the loss gradient. 

\paragraph{Task-Specific Adversarial Losses} For each task $t$, where $t\neq{j}$, its task-specific representations are passed as adversarial representations to the output head of task $j$ to ensure that they do not carry information related to other tasks. Thus, the total task-specific adversarial loss ($\mathcal{L}_{Adv}^j$) of task $j$ is defined as follows:
\begin{gather}
    \mathcal{L}_{Adv}^j = \sum_{t}{\mathcal{L}}_{Adv}^t\,\,\,\,\forall{t}\in{\mathbb{T} - \{j\}},
\end{gather}
During backpropagation, the adversarial loss gradients are reversed by the gradient reversal layer when updating the corresponding task-specific extractors so that the weights are updated in the ascending direction along the gradients.

Finally, combining the three main loss terms, the overall loss for task $j$ can be defined as follows:
\begin{gather}
    \mathcal{L}^j = \mathcal{L}^j_O - \alpha_{1}\mathcal{L}^j_D -\alpha_{2}\mathcal{L}^j_{Adv},
    \vspace{-25pt}
\end{gather}
where $\alpha_{1}$ and $\alpha_{2}$ are hyper-parameters determining the importance of the adversarial debiasing loss terms.
\begin{table*}[h!]
   \centering
   \footnotesize
   \scalebox{0.9}{\begin{tabular}{c|l|llll|l}
        \hline \hline
        & & \multicolumn{4}{c|}{\textbf{Validation}} & \textbf{Test} \\ \hline
        Code & Model(s) & Emo-CCC & Cou-UAR & Age-MAE & $S_{MTL}$ & $S_{MTL}$\\
        \hline
        Baseline & ComParE & 0.416 & 0.506 & 4.222 & 0.349 & 0.335\\
        \hline
        V1 & Burst2Vec WL$^\spadesuit$ & 0.653 & 0.672 & 3.672 & 0.448 & 0.421\\
        V2 & \,\,-- Adv & 0.645$^\clubsuit$ & 0.676$^\clubsuit$ & 3.709$^\clubsuit$ & 0.445 & -\\
        \hline
         & Burst2Vec HB & & & & &\\
        V3 & \,\,--Adv & 0.652 & 0.606 & 3.594 & 0.443 & -\\
        V4 & Burst2Vec HL & 0.675 & 0.638 & 3.638 & 0.449 & -\\
        V5 & \,\,--Adv & 0.675$^\diamondsuit$ & 0.631$^\diamondsuit$ & 3.694$^\diamondsuit$ & 0.444 & -\\
        V6 & Burst2Vec HL$^\spadesuit$ & 0.676$^\diamondsuit$ & 0.644 & 3.640$^\diamondsuit$ & 0.450 & 0.417\\
        V7 & \,\,+Oversampling$^\spadesuit$ & 0.669$^{\diamondsuit\heartsuit}$ & 0.630 & 4.233$^{\diamondsuit\heartsuit}$ & 0.410 & -\\
        \hline
        Ensemble A & V1 -- V3, V6  & 0.704$^{\clubsuit\diamondsuit\heartsuit}$ & \textbf{0.690}$^{\clubsuit\diamondsuit\heartsuit}$ & \textbf{3.580}$^{\clubsuit\diamondsuit\heartsuit}$ & \textbf{0.465} & \textbf{0.435}\\
        Ensemble B & V1 -- V3, V5, V6 & 0.708$^{\clubsuit\diamondsuit\heartsuit}$ & 0.688$^{\clubsuit\diamondsuit\heartsuit}$ & 3.589$^{\clubsuit\diamondsuit\heartsuit}$ & 0.465 & 0.432\\
        Ensemble C & V1 -- V6 & \textbf{0.711}$^{\clubsuit\diamondsuit\heartsuit}$ & 0.682$^{\clubsuit\diamondsuit\heartsuit}$ & 3.586 & 0.464 & 0.432\\
        \hline\hline
    \end{tabular}}
  \caption{Results on the validation and test sets of the ExVo challenge dataset. $^\spadesuit$ indicates the models trained on our pre-processed version of the audio files. $^\clubsuit$, $^\diamondsuit$ and $^\heartsuit$ indicate statistically significant changes (p-value $\leq 0.05$) as compared to Burst2Vec WL$^\spadesuit$, Burst2Vec HL and Burst2Vec HL$^\spadesuit$, respectively. Note that the statistical significance tests are carried out for each task ($t$-test for the emotion and age tasks and Cochran's Q test for the country task) and not for the harmonic mean values ($S_{MTL}$). The best results are highlighted in bold.}
  \label{Tab:model_results}
\end{table*}

\section{Experimental Settings}
\subsection{Data Pre-processing}
Since there is some distortion in the provided processed audio files, we pre-processed the raw files. To be comparable with the baseline methods, the raw audio files were re-sampled to $16$\,kHz to for faster processing~\cite{ren2020generating}, converted to mono, and normalised to $-3$ decibels with the RMS method. Moreover, to ensure effectiveness during training, we perform 0-1 normalisation on the age labels. As mentioned in \S\ref{subsec:data_analysis}, there is an apparent dataset imbalance in age and country labels. Therefore, we explore oversampling samples belonging to the minority classes or label ranges via random sampling. Due to the multi-task setting, we perform oversampling based on the cross-label distribution of the three tasks, preventing scenarios where oversampling improves the label balance on one task while worsening the balance of other tasks.

\subsection{Model Variants} We explore three versions of the Burst2Vec model with different speech representation models, all initially fine-tuned on the 960\,hours of the Librispeech~\cite{librispeech}: Burst2Vec WL using the wav2vec 2.0 Large model~\footnote{https://huggingface.co/facebook/wav2vec2-large-960h}, Burst2Vec HB using the HuBERT Base model~\footnote{https://huggingface.co/facebook/hubert-base-ls960}, and Burst2Vec HL using the HuBERT Large model~\footnote{https://huggingface.co/facebook/hubert-large-ls960-ft}.


We also experiment with variants of the above models without adversarial debiasing (--Adv) and using the oversampled dataset (+Oversampling) to assess their individual contribution to model performance. 
See Appendix~\ref{sec:train_configs} for explanation of the training configurations and hyperparameters.

\paragraph{Model Ensembling} We perform model ensembling by averaging the predicted emotion intensities and age from each model. For the country task, we average the softmax probabilities from each model and take the country with the highest mean probability as the final prediction.

%

\section{Results}

Following the Challenge's white paper~\cite{BairdExVo2022}, we report the CCC score for the emotion task (Emo-CCC), the Unweighted Average Recall for the country task (Cou-UAR), the MAE value for the age task (Age-MAE), and the harmonic mean ($S_{MTL}$) of the three task scores (main metric). We also perform statistical significance tests with independent two-sample $t$-test~\cite{student1908probable} for the emotion and age tasks and Cochran's Q test~\cite{cochran1950comparison} method for the country task. Results are reported in Table~\ref{Tab:model_results}, with the `Test' column populated with our five official submissions and the highest performing official baseline.

\paragraph{Effect of Pre-trained Speech Representation Models} Compared to all baseline models, all our model variants perform much better, with Burst2Vec HL$^\spadesuit$ (V6) performing the best on the validation set. We believe this large performance gain is  due to the ability of pre-trained models to capture richer information from the input audio signals.

\paragraph{Different Speech Representation Models} Comparing the three main variants without adversarial debiasing (V2, V3, and V5), their overall performances on the validation set are on-par with each other. However, each non-adversarial sub-variant best performs on one particular task (i.\,e., V2, V3, and V5 are best for the country, age, and emotion task, respectively). This can be because, without adversarial debiasing, the model would be biased by capturing features more useful towards certain tasks, and this depends on the architecture of the speech representation model.

\paragraph{Effect of Adversarial Debiasing} 
From pair-wise comparisons between sub-variants of Burst2Vec WL (V1 and V2) and Burst2Vec HL (V4 and V5), with and without adversarial debiasing, the overall performance increases with adversarial debiasing. At the task level,  debiasing helps improve performance on tasks that the non-adversarial sub-variant performs worse upon.
For instance, for Burst2Vec HL, the adversarial variant (V4) performs better than the non-adversarial counterpart (V5) on the country (+ 0.01 CCC) and age tasks (- 0.06 MAE) and performs on-par on the emotion task, which is the best-performing task for V5.

\paragraph{Effect of Oversampling} 
Comparing V6 and V7, we observe that oversampling drastically worsens the performance on all tasks, which is related to the severe dataset imbalance. As samples in the minority region have to be sampled multiple times to balance the distribution, the model is misguided to learn false relationships in the dataset. 

\paragraph{Effect of Using Different Audio Files} Between V4 and V6, there is a small improvement overall (+\,0.001 $S_{MTL}$) when training the model with our version of the audio files, due to the distortion present in the provided version.

\paragraph{Effect of Model Ensembling} As expected, model ensembling improves the overall performance on both the validation and test sets by a large margin. In particular, Ensemble A yields the best performance on both sets. We posit that this reduces uncertainty in the model predictions. 

\section{Conclusions}
We presented our winning approach, Burst2Vec -- an adversarial multi-task approach for jointly predicting emotion, age, and (country of) origin from vocal bursts. We incorporated pre-trained speech representation models to capture both acoustic 
and linguistic 
information from raw waveforms. We also introduced the concept of model debiasing via adversarial training to disentangle each input representation into shared and task-specific representations.

Our variants perform remarkably better than the baselines and score the highest in the Challenge leaderboard. 
Namely, Burst2Vec WL$^\spadesuit$ achieves the best single-model performance, while ensembling further improves these results. In future works, one could explore model architectures which combine pre-extracted features and features from a speech representation model as well as experimenting with different debiasing techniques for multi-task learning.  

\bibliography{example_paper}
\bibliographystyle{icml2022}

\newpage
\appendix
\onecolumn
\section{Appendix -- wav2vec 2.0} \label{sec:wav2vec2_description}
The wav2vec 2.0~\cite{wav2vec2} model is a self-supervised framework learning the representations from raw audio signals. It contains a local feature encoder, which takes the raw audio samples as input into several convolutional blocks and outputs $768$-dimensional (base one) or $1024$-dimensional embeddings (large one). Afterwards, the embeddings are discretised to a finite set of local representations by Gumbel softmax~\cite{gruloss} in the quantisation module. Further, Transformers~\cite{transformer} are applied for contextualised representations. 


\section{Appendix -- Hidden-Unit BERT (HuBERT)} \label{sec:hubert_description}
HuBERT~\cite{hubert} is another approach for self-supervised speech representation learning, utilising an offline k-means clustering step to provide labels for the prediction processes. Specifically, with the CNN encoder, continuous speech features are generated from the raw audio. Afterwards, in order to capture long-range temporal relations, the features are randomly masked and fed into the BERT encoder to predict pre-determined cluster assignments. The prediction loss is calculated based only on the masked regions, which enables the HuBERT model to learn both acoustic and language models from the continuous inputs~\cite{hubert}. 


\section{Appendix -- Training Configurations and Hyperparameters} \label{sec:train_configs}
For all experiments, $\alpha_{1}$ and $\alpha_{2}$ are set to 0.5 and 0.25, respectively. We also use the Adam optimiser~\cite{adam} and set the learning rate to $10^{-5}$ with the batch size of 32 (See Table~\ref{Tab:train_hyper} for the full list of hyperparameters). During training, we configure each batch to correspond to one of the three tasks. Also, we apply zero padding to the input audio files in each batch in order to unify the input lengths. On a single NVIDIA RTX A6000 GPU, it takes around 24 hours up to 36 hours to train a model, depending on the dataset size.

\begin{table*}[h!]
   \centering
   \footnotesize
   \begin{tabular}{l|l}
        \hline\hline
        \textbf{Hyperparameter} & \textbf{Value}\\
        \hline
        \textit{\textbf{General}} & \\
        Batch Size & 32\\
        Max Epochs & 100\\
        Validation Frequency (Iterations) & 800\\
        Early Stopping Patience (Number of Validation Steps) & 20\\
        Early Stopping Metric & CCC Metric of the Emotion Task\\
        \hline
        \textit{\textbf{Optimiser}} & \\
        Optimiser Name & Adam\\
        Learning Rate & $10^{-5}$\\
        \hline
        \textit{\textbf{Loss}} & \\
        $\alpha_{1}$ & 0.5\\
        $\alpha_{2}$ & 0.25\\
        \hline\hline
    \end{tabular}
  \caption{Training hyperparameters used for all model variants experimented in this project.}
  \label{Tab:train_hyper}
\end{table*}


\end{document}